\documentclass[a4paper,11pt]{article}
\pdfoutput=1 
\usepackage{jheppub}
\usepackage{subfig}

\title{Final results of the search for \boldmath $\nu_{\mu} \to \nu_e$ oscillations with the OPERA detector in the CNGS beam}
\collaboration{The OPERA collaboration}
\author[a]{N. Agafonova,}
\author[b]{A. Aleksandrov,}
\author[c]{A. Anokhina,}
\author[d]{S. Aoki,}
\author[e]{A. Ariga,}
\author[e,f]{T. Ariga,}
\author[g]{A. Bertolin,}
\author[h]{C. Bozza,}
\author[g,i]{R. Brugnera,}
\author[b,j]{A. Buonaura,}
% ...

\affiliation[a]{INR - Institute for Nuclear Research of the Russian Academy of Sciences,\\RUS-117312 Moscow, Russia}
\affiliation[b]{INFN Sezione di Napoli,\\I-80125 Napoli, Italy}
\affiliation[c]{SINP MSU - Skobeltsyn Institute of Nuclear Physics, Lomonosov Moscow State University,\\RUS-119991 Moscow, Russia}
\affiliation[d]{Kobe University,\\J-657-8501 Kobe, Japan}
\affiliation[e]{Albert Einstein Center for Fundamental Physics, Laboratory for High Energy Physics (LHEP), University of Bern,\\ CH-3012 Bern, Switzerland}
\affiliation[f]{Faculty of Arts and Science, Kyushu University,\\ J-819-0395 Fukuoka, Japan}
\affiliation[g]{INFN Sezione di Padova,\\ I-35131 Padova, Italy}
\affiliation[h]{Dipartimento di Fisica dell'Universit\`a di Salerno and ``Gruppo Collegato''  INFN,\\ I-84084 Fisciano (Salerno), Italy}
\affiliation[i]{Dipartimento di Fisica e Astronomia dell'Universit\`a di Padova,\\ I-35131 Padova, Italy}
\affiliation[j]{Dipartimento di Fisica dell'Universit\`a Federico II di Napoli,\\ I-80125 Napoli, Italy}
% ...

\emailAdd{matteo.tenti@bo.infn.it}
\emailAdd{zemskova@numail.jinr.ru}

\abstract{The OPERA experiment has discovered the tau neutrino appearance in the CNGS muon neutrino beam, in agreement with the 3 neutrino flavour oscillation hypothesis. 
The OPERA neutrino interaction target, made of Emulsion Cloud Chambers, was particularly efficient in the reconstruction of electromagnetic showers. Moreover, thanks to the very high granularity of the emulsion films, showers induced by electrons can be distinguished from those induced by $\pi^0$s, thus  allowing the detection of charged current interactions of electron neutrinos. In this paper the results of the search for electron neutrino events using the full dataset are reported. An improved method for the electron neutrino energy estimation is exploited.
Data are compatible with the 3 neutrino flavour mixing model expectations and are used to set limits on the oscillation parameters of the 3+1 neutrino mixing model, in which an additional mass eigenstate $m_{4}$ is introduced. At high $\Delta m^{2}_{41}$ $( \gtrsim 0.1~\textrm{eV}^{2})$, an upper limit on $\sin^2 2\theta_{\mu e}$ is set to 0.021 at 90\% C.L. and $\Delta m^2_{41} \gtrsim 4 \times 10^{-3}~\textrm{eV}^{2}$  is excluded for maximal mixing in appearance mode.}
\keywords{Neutrino Detectors and Telescopes (experiments), Oscillation}
\arxivnumber{1803.11400}

\begin{document}
\maketitle
\flushbottom
\section*{Introduction }
\addcontentsline{toc}{section}{Introduction}
\label{sec:intro}

\par The main goal of the OPERA experiment was the search for $\nu_{\mu} \to \nu_{\tau}$ oscillations in the region of squared mass difference pointed out by experiments on the atmospheric neutrinos through the observation of the tau neutrino appearance in a muon neutrino beam~\cite{OPERAproposal}. 
The OPERA apparatus~\cite{OPERA2009} located at the Gran Sasso Underground Laboratory was exposed from 2008 to 2012 to the CERN Neutrinos to Gran Sasso (CNGS) muon neutrino beam~\cite{CNGSbeam1}, produced at a distance of about 730~km. 
The analysis of a data set released in 2015 led to the identification of five tau neutrino candidates with an expected background of 0.25, thus excluding the no-oscillation hypothesis with a significance greater than 5$\sigma$~\cite{tau5Sigma}. 
\par The OPERA nuclear emulsion target also allows the identification of $\nu_{e}$ (or $\bar{\nu}_{e}$) charged current (CC) interactions.
The outcome of the search for electron neutrinos in the data collected in 2008--2009 was reported in~\cite{OPERAnue}. 
Here, we report the results of an improved analysis on the complete dataset, corresponding to 17.97 $\times$ 10$^{19}$~p.o.t., which represents an increase of the exposure by a factor 3.4 with respect to the previous analysis. 
\par Data compared with the expectation from the 3 neutrino flavour oscillation model allow  setting an upper limit on the $\nu_{e}$ appearance probability. Moreover, the hypothesis of a sterile neutrino, as hinted by LSND~\cite{LSND},  MiniBooNE~\cite{MiniBooNE}, reactor~\cite{reactoranomaly}, and radio-chemical~\cite{GALLEX,SAGE} experiments, is tested. Limits on $\Delta m^{2}_{41}$ and $\sin^{2}2\theta_{\mu e} = 4|U_{e4}|^{2}|U_{\mu4}|^{2}$ are derived.
\section{The OPERA experiment}
\label{sec:experiment}

\par The detector was composed of two identical super modules, each consisting of a target section and a magnetic iron spectrometer. Each target section contained about 75000 Emulsion Cloud Chamber~\cite{Niu:1971xu} modules, hereafter called bricks, for a total target mass of about 1.25~kt. The bricks were arranged in vertical walls interleaved with planes of horizontal and vertical scintillator strips, which formed the Target Tracker (TT). Each brick consisted of 56 1-mm lead plates interspersed with 57 emulsion films. It had a section of $12.7 \times 10.2$~cm$^2$ and a thickness of 7.5~cm, which corresponds to $\sim$10 radiation lengths ($X_0$). A pair of emulsion films (Changeable Sheet or CS doublet) were packed and glued externally to the downstream face of each brick. The CS doublet acted as an interface between the cm-resolution TT and $\mu$m-resolution emulsion, thus triggering the film development. The spectrometer, instrumented with planes of Resistive Plate Chambers (RPC) and drift tube stations, provided the charge and momentum measurements of muons~\cite{OPERAspec}. 

\par The CNGS beam was an almost pure muon neutrino beam having a mean energy of 17~GeV. The contaminations, in terms of CC interactions, of $\nu_{e}$, $\bar{\nu}_{e}$ and $\bar{\nu}_{\mu}$ were 0.88\%, 0.05\% and 2.1\%, respectively.

\par The TT hits of an event occurring on time with the CNGS beam were used to track charged particles in the target region and to provide a calorimetric energy measurement.  
The TT information was exploited to rank the bricks according to their probability to contain the neutrino interaction vertex~\cite{OPERA_BF}.  
The highest probability brick was extracted from the detector. The attached CS doublet was scanned and the pattern of reconstructed tracks would either confirm the prediction of the electronic detector or act as veto and thus trigger the extraction of neighbouring bricks. In case of positive outcome, the brick emulsion films were developed and analysed. 
CS tracks were then matched with those reconstructed in the downstream films of the brick. These latter tracks were then followed upstream in the brick to their origins. Around the most upstream track disappearance point, a 1~cm$^2$ wide area in 10 downstream and 5 upstream emulsion films was scanned. All tracks and vertices in this volume were reconstructed, and short-lived particle decays were searched for.
In case no neutrino interaction was found in the brick, the search would continue in less probable bricks. More details on the search and reconstruction of neutrino events in the bricks are given in~\cite{OPERA1stTau,OPERAds,OPERA3rdTau}.

\section{Search for $\nu_{e}$ candidates}
\label{sec:nuesearch}

\par The brick acted as a high sampling calorimeter with more than five active layers every $X_{0}$ over a total thickness of 10~$X_{0}$. For most of $\nu_e$ CC interactions, the path of the electron in the brick is long enough for the electromagnetic (e.m.) shower to develop, thus allowing its detection and reconstruction in the emulsion films. On the other hand, the size of the standard scanned volume along the beam direction corresponds to about 1.8~$X_0$, which is too short for the e.m.~shower to develop.
The search for electrons is therefore performed in an extended scanned volume applying a dedicated procedure to the 0$\mu$ tagged events, i.e. events having no reconstructed three-dimensional muon track and less than 20 fired TT/RPC planes.
A search is performed in the CS doublet for track segments less than 2~mm apart from the extrapolation point of each track originated from the interaction vertex (primary tracks). Moreover, the direction of candidate CS tracks is required to be compatible within 150~mrad with that of the primary track. If at least three tracks are found, additional scanning along the primary track is performed aiming at the detection of an e.m.~shower~\cite{OPERAnue}.

\par Detected showers are carefully inspected, by visual scan, in the first two films downstream the interaction vertex to assess whether they are produced by a single particle. Thanks to the high granularity of the OPERA  nuclear emulsions, one can recognize an $e$-pair from $\gamma$ conversion when the $e$-pair tracks are separated by more than 1 $\mu$m. 
Once the origin of the e.m.~shower is confirmed as due to a single charged particle, the event is classified as $\nu_{e}$ candidate. 
A classifier algorithm~\cite{OPCARAC}, is applied to select neutrino interaction events fully contained in the fiducial volume of the detector.

\par
\begin{figure}[t] \centering
\includegraphics[width=0.9\textwidth]{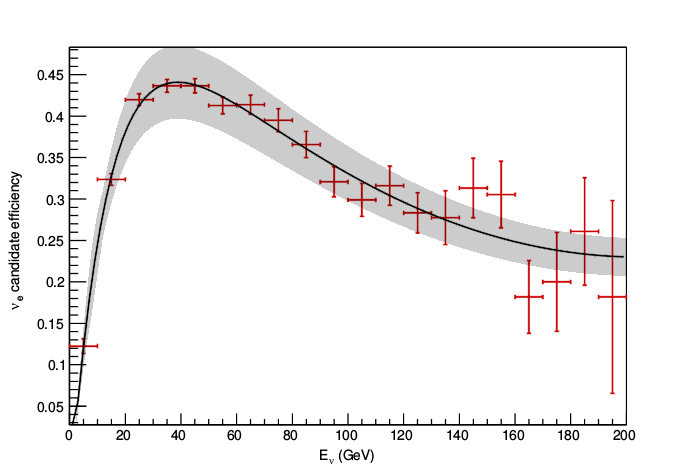}
\caption{\label{fig:nue_eff} The $\nu_{e}$ CC candidate selection efficiency as a function of the neutrino energy. Error bars indicate the MC statistical errors. The black line is a fitted curve to the simulated efficiency points and the grey area represents the systematic error.}  
\end{figure}

 The efficiency of the procedure applied to search for $\nu_{e}$ CC interactions, shown in figure~\ref{fig:nue_eff},  is computed by Monte Carlo (MC) simulation using the standard OPERA simulation framework~\cite{OPERA3rdTau}. Neutrino fluxes are determined by FLUKA~\cite{Fluka1, Fluka2} based simulation of the CNGS beamline~\cite{CNGSbeam2}.
Neutrino interactions in the target are generated using the GENIE v2.8.6 generator~\cite{GENIE1, GENIE2}.
The simulated efficiency points are then fitted with an empirical function. Since the number of hit TT/RPC planes and presence of long tracks in neutrino events are correlated with the neutrino energy, as a result of 0$\mu$ tagging criterion, a drop in the efficiency is expected at high energies.   
\par In 2008--2012, the OPERA detector had collected data corresponding to $17.97 \times 10^{19}$~p.o.t. In total, 19505 on-time events have been registered in the target volume. Out of them, 5868 events have a reconstructed neutrino interaction vertex in the first or second most probable brick and 1281 are tagged as 0$\mu$ events. The 0$\mu$ sample is further reduced to 1185 events by exclusion of not contained ones. In this sample the number of $\nu_e$ candidates is 35.

\par The number, $N_{beam}$, of $\nu_{e}$ candidates from CC interactions of $\nu_e$ and $\bar{\nu}_e$ beam components is estimated using a data-driven approach from the number of observed events with no charged leptons, 0$l$, which are 0$\mu$ events not identified as $\nu_e$ candidates, $n_{0l} = 1185-35$, according to: 
 
\begin{equation}
N_{beam} =n_{0l}\frac{R^{CC}_{\nu_{e}} \big \langle \varepsilon_{CC}^{\small{\nu_{e}}} (\nu_{e})\big \rangle}{ \sum\limits_{i=\mu , e} \sum\limits_{j=CC, NC} R_{\nu_{i}}^{j} \big \langle \varepsilon_{j}^{0l} (\nu_{i})\big \rangle},
\label{norm}
\end{equation}

\noindent where $\big \langle \varepsilon_{j }^{\small{\nu_{e}}} (\nu_{i})\big \rangle$ and $\big \langle \varepsilon_{j}^{0l} (\nu_{i})\big \rangle$ are the efficiencies for the $\nu_{i}$ interactions ($i$ = $\mu$, $e$), convoluted with the CNGS $\nu_{i}$ flux, $\phi_{\nu_{i}}$, and its $j$-type interaction cross sections, $\sigma_{\nu_i}^{j}$ ($j$ = CC or NC), to be reconstructed as a $\nu_{e}$ candidate or as a 0$l$ event, respectively, i.e.:
\begin{equation}
\left. \big \langle \varepsilon_{j }^{\small{\nu_{e}} (0l)} (\nu_{i})\big \rangle = \int \phi_{\nu_{i}} \varepsilon_{j}^{\small{\nu_{e}} (0l)} \sigma_{\nu_i}^{j}~dE \middle/ \int \phi_{\nu_{i}} \sigma_{\nu_i}^{j}~dE, \right.
\label{effdef} 
\end{equation}
while $R_{\nu_{i}}^{j}$ are the interaction rates of neutrino and antineutrino:
\begin{equation}
R_{\nu_{i}}^{j} = \int \phi_{\nu_{i}} \sigma_{\nu_i}^{j}~dE.
\label{ratedef} 
\end{equation}

\noindent The expected rates $R_{\nu_{i}}^{j}$, for $10^{19}$~p.o.t. and 1~kt target mass, and the efficiencies $\langle \varepsilon_{CC }^{0l} (\nu_{i})\big \rangle$ are reported in table~\ref{rates}. The efficiency $\langle \varepsilon_{CC }^{\small{\nu_{e}}} (\nu_{e})\big \rangle$ is 0.375. 

\begin{table}[t]
\centering
\begin{tabular}{|c|rc|rc|}
\cline{2-5}
\multicolumn{1}{c}{} & \multicolumn{4}{|c|}{$R_{\nu_{i}}^{j}~~\big( \big \langle \varepsilon_{j }^{0l} (\nu_{i})\big \rangle \big)$} \\ 
\cline{2-5}
\multicolumn{1}{c}{} & \multicolumn{2}{|c|}{$j = CC$} & \multicolumn{2}{c|}{$j = NC$} \\ \hline
$i = \mu$      &  640.4 & (0.029)  &  215.2 & (0.296)  \\ \hline
$i = e$        &  5.9 & (0.078)  &   2.2 & (0.307) \\ \hline
\end{tabular}
\caption{Expected rates of neutrino and antineutrino of type $i$ interacting by the $j$ process normalized to 10$^{19}$~p.o.t. and 1~kt target mass. In brackets the efficiencies of $\nu_{i}$ interactions, convoluted with the CNGS $\nu_{i}$ flux and its $j$-type interaction cross sections, to be reconstructed as a 0$l$ event.}
\label{rates}
\end{table}

The combined systematic error from flux normalization, $\nu_{e}$/$\nu_{\mu}$ flux ratio, detection efficiencies and cross section uncertainties is conservatively estimated as 20\%  and 10\% for neutrino energies  below and above 10~GeV, respectively. The expected value of $N_{beam}$ amounts to 30.7 $\pm$ 0.9 (stat.) $\pm$ 3.1 (syst.) events. The expected numbers obtained with the above mentioned procedure are insensitive to systematic effects on the efficiencies up to the location level being common to $\nu_{e}$ and $0l$ events.

\par Background contributions are: $\nu_{\tau}$ CC interactions with $\tau \to e$ decays, $0\mu$ events with $\pi^0 \to \gamma \gamma$ decay with prompt $\gamma$ conversion, and the electron-positron pair misidentified as an electron. The first type of background arises when the $\tau$-lepton and its daughter electron track cannot be distinguished. It is estimated by MC simulation assuming the 3-flavour $\nu_{\mu} \to \nu_{\tau}$ oscillation scheme and oscillation parameters from~\cite{PDG2016}. It amounts to 0.7 $\pm$ 0.2 (syst.) events.

\par The background due to $\pi^{0}$ is derived from the data by counting the number of events fulfilling the criteria for a $\nu_{e}$ candidate with a $\gamma$ converting in the second or third lead plate downstream of the interaction vertex. That number is converted into the probability to observe background $\nu_{e}$ candidates due to $\gamma$ conversions in the first lead plate, taking into account the radiation length~\cite{OPERAnue}. We expect 0.5 $\pm$ 0.5 (stat.) such events. 

\par Summarizing, the expected number of background events, $N_{bkg}$, amounts to 1.2 $\pm$ 0.5 (stat.) $\pm$ 0.2 (syst.), while the sum of expected events from $\nu_e$ and $\bar{\nu}_e$ beam components and backgrounds is 31.9 $\pm$ 1.0 (stat.) $\pm$ 3.1 (syst.).

\section{Energy reconstruction}
\label{sec:erec}

\par
\begin{figure}[t]
 \centering
\includegraphics[width=0.7\textwidth]{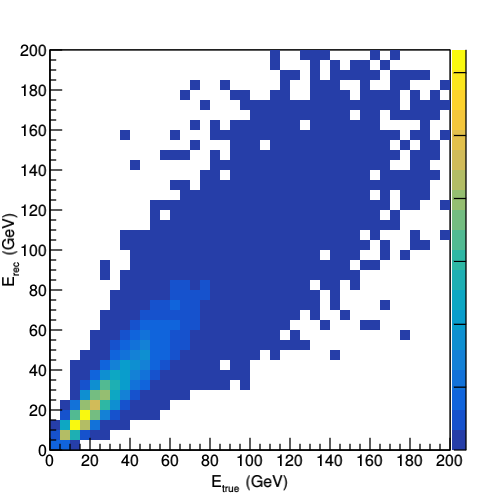}
\caption{\label{fig:Erec}
Reconstructed versus MC energy for $\nu_e$ CC events in the OPERA Target Trackers.}
\end{figure}

 The longitudinal and transverse development of e.m.~showers are well parametrised both for homogeneous and sampling calorimeters~\cite{EM_Longo1975,EM_Param2000,Leroy2000}, hence the energy shape analysis can be used to improve the electron energy reconstruction. 
The e.m.~shower shape parametrisation together with the knowledge of the interaction vertex position and of the direction of the electron, obtained from the emulsion data, are the basis of the method applied in this analysis, as detailed in~\cite{ErecProceeding}. In this approach, the TT signals from e.m.~shower, initiated by the electron at the interaction vertex, and the hadronic shower are separated. The hadronic shower energy is estimated by the calorimetric measurement in the TT. The electron energy is assessed by the shape analysis of the e.m.~shower profiles. The reconstructed energy of $\nu_e$ CC events versus their true energy is shown in figure~\ref{fig:Erec}. The accuracy of the energy reconstruction of $\nu_{e}$ CC events can be parametrized as:

\begin{equation}
\frac{\Delta E}{E} = 0.18 + \frac{0.55}{\sqrt{E(\textrm{GeV})}} 
\end{equation}

\noindent which is improved with respect to previous analysis~\cite{OPERAnue}.

\section{Neutrino oscillation analysis}
\label{analysis}

\subsection{The 3 neutrino flavour mixing model}
\label{sec:stdosc}

\par The neutrino oscillation phenomenon can lead to a non null difference, $n_{osc}$, between the number of observed $\nu_{e}$ candidates, $n_{obs}$, and that expected from the beam contamination, $N_{beam}$ and from background, $N_{bkg}$, i.e.

\begin{equation}
n_{osc} = n_{obs} - N_{beam} - N_{bkg}.
\label{nosc}
\end{equation}
\noindent On the other end, assuming only contributions from $\nu_{\mu} (\bar{\nu}_{\mu}) \rightarrow \nu_{e}  (\bar{\nu}_{e})$ oscillations the expected value of $\nu_{e}$ candidates from oscillations, $N_{osc}$, is:

\begin{equation}
N_{osc} = k \cdot \int \phi_{\nu_{\mu}} P_{\mu e} \sigma_{\nu_e} \varepsilon_{\nu_{e}} dE, 
\label{neosc}
\end{equation}
where $k$ is the averaged number of nuclei over the data taking period, $P_{\mu e}$ is the oscillation probability, $\varepsilon_{\nu_{e}}$ and $\sigma_{\nu_e}$ are the $\nu_{e}$ detection efficiency and cross section, respectively, $\phi_{\nu_{\mu}}$ is the $\nu_{\mu}$ integrated flux and $E$ is the neutrino energy. The average appearance probability, $\left<P_{\mu e}\right>$, can be derived as:

\begin{equation}
\left<P_{\mu e}\right> = \frac{n_{osc}} {k \cdot \int \phi_{\nu_{\mu}} \sigma_{\nu_{e}} \varepsilon_{\nu_{e}} dE}.
\label{posc}
\end{equation}
An optimal interval on neutrino energy, 0--40~GeV, was introduced in order to maximize the sensitivity on the average appearance probability. $\left<P_{\mu e}\right>$ is model-independent, and energy interval defined above can be used for any oscillation analysis of the OPERA data not based on the energy shape analysis. The 90\% C.L. upper limit, obtained using the Feldman and Cousins method (F\&C)~\cite{FC}, is $\left<P_{\mu e}\right> < 3.5 \times 10^{-3}$ for neutrino energies $0 < E_{\nu} < 40~\textrm{GeV}$. A similar result, $\left<P_{\mu e}\right> < 3.7 \times 10^{-3}$, is obtained using the Bayesian technique~\cite{Bayesian}.
 
\par Constraints on the oscillation parameters are derived within the framework of the 3 neutrino flavour oscillation model. The expected number of $\nu_{e}$ candidates ($N_{3\nu}^{exp}$) depends on the disappearance of $\nu_{e}$ ($\bar{\nu}_{e}$) beam contamination and on $\nu_{e}$ ($\bar{\nu}_{e}$) appearance from $\nu_{\mu}$ ($\bar{\nu}_{\mu}$) beam components. Both these contributions are taken into account. The oscillation probabilities are evaluated from:

\begin{equation}
\begin{split}
P(\nu_{\alpha}  \rightarrow \nu_{\beta}) = & \delta_{\alpha \beta}~- \\
& - 4 \sum_{i>j}(U_{\alpha i}U_{\beta i} U_{\alpha j} U_{\beta j}) \sin^2 \left(1.27~\Delta m_{ij}^2\left(	\textrm{eV}^{2}/c^{4}\right) \frac{L(\textrm{km})}{E(\textrm{GeV})} \right).
\end{split}
\end{equation}

\par
\begin{figure}[!tbp]
  \centering
  \subfloat[]{\includegraphics[width=0.5\textwidth]{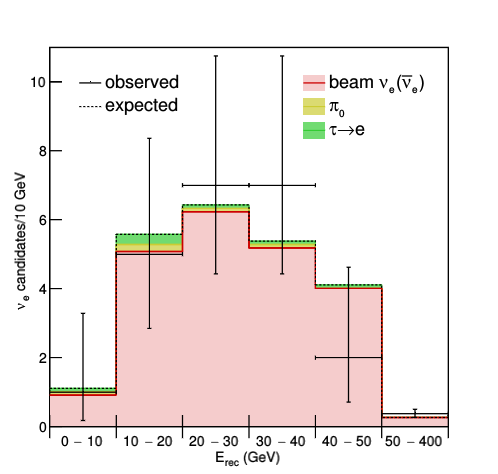}\label{fig:energy_spectra_no_osc}}
  \subfloat[]{\includegraphics[width=0.5\textwidth]{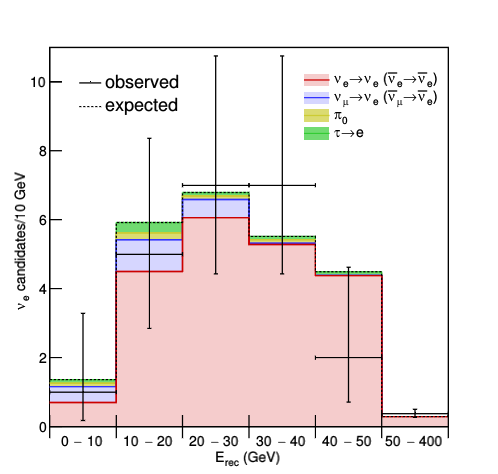}\label{fig:energy_spectra_osc}}
  \caption{The reconstructed energy distributions of the observed $\nu_{e}$ candidates, the expected background and beam $\nu_{e}$ and $\bar{\nu}_{e}$ components: (a) assuming no oscillations; (b) in case of 3 neutrino flavour mixing with the parameters from~\cite{PDG2016}.}
\end{figure}

 Using the standard parametrization of the mixing matrix $U$, the values of $\theta_{13}$, $\theta_{23}$, $\delta_{CP}$ and $\Delta m^{2}_{ij}$ from~\cite{PDG2016}, the expected number of $\nu_{e}$ candidates, including 1.2 background events, is $N_{3\nu}^{exp} =$ 34.3 $\pm$ 1.0 (stat.) $\pm$ 3.4 (syst.). Matter effects are taken into account, more details are given in section~\ref{sec:3plus1}. $N_{3\nu}^{exp}$ is consistent with 35 observed $\nu_{e}$ candidate events. Figure~\ref{fig:energy_spectra_no_osc} shows the energy distributions of observed events, of the expectation from $\nu_{e}$ and $\bar{\nu}_{e}$ beam components, assuming no oscillations, and of background events ($\pi^{0}$ and $\tau \rightarrow e$). In figure~\ref{fig:energy_spectra_osc} the energy distribution of observed events is compared with the expectation from $\nu_e(\bar{\nu}_e) \to \nu_e(\bar{\nu}_e)$ and $\nu_{\mu}(\bar{\nu}_{\mu}) \to \nu_e(\bar{\nu}_e)$ oscillations channels, and from background components. The background components, $\pi_0$ and $\tau \to e$, are assumed to be unaffected by oscillations. In the optimized neutrino energy range (0--40~GeV), the 90\% C.L. upper limit on $\sin^2 2\theta_{13}$ is 0.43.

\subsection{Analysis in the 3+1 neutrino mixing model}
\label{sec:3plus1}

\par
\begin{figure}[t]
 \centering
\includegraphics[width=0.8\textwidth]{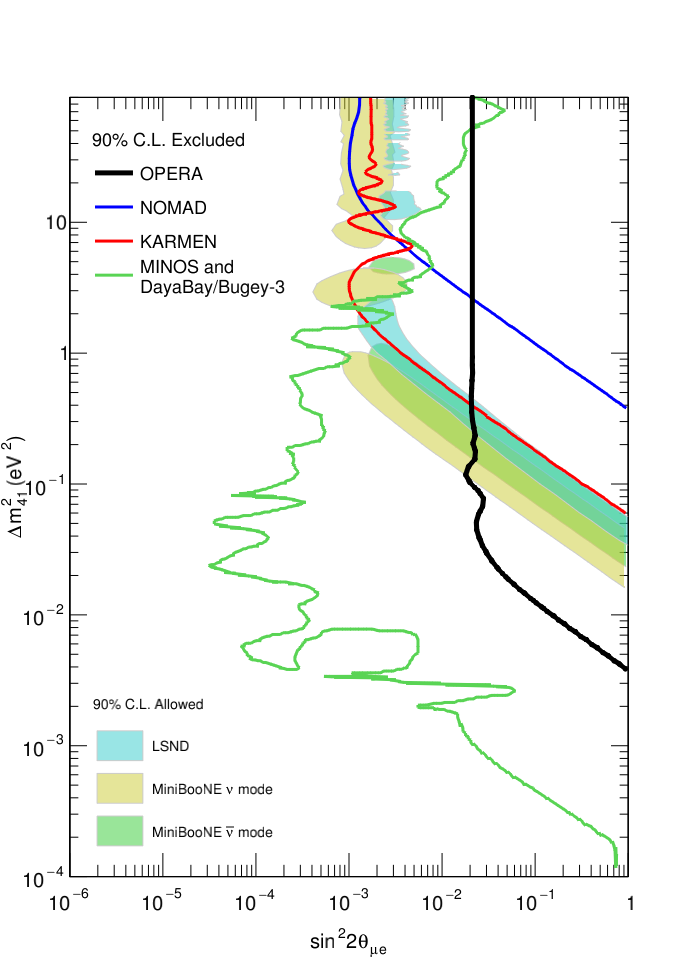}
\caption{\label{fig:exclusion_plot}
The 90\% C.L. exclusion plot in the $\Delta m^{2}_{41}$ and $\sin^{2}2\theta_{\mu e}$ plane is shown (black line) together with the 90\% C.L. allowed region obtained by LSND (cyan) and MiniBooNE (yellow and green for $\nu$ and $\bar{\nu}$ mode respectively). The blue, red and green lines represent the 90\% C.L. exclusion regions obtained by NOMAD~\cite{NOMAD}, KARMEN~\cite{KARMEN} and the MINOS and DayaBay/Bugey-3 joint analysis~\cite{MINOSDB}, respectively.} 
\end{figure}

 The excess of $\nu_{e}$ and $\bar{\nu}_{e}$ reported by the LSND~\cite{LSND}, MiniBooNE~\cite{MiniBooNE}, reactor~\cite{reactoranomaly} and radio-chemical~\cite{GALLEX,SAGE} experiments may be interpreted as due to the presence of light, $\mathcal{O}(1~\textrm{eV/c$^{2}$})$, sterile neutrinos. OPERA can test the hypothesis of the presence of a sterile neutrino by comparing the observed $\nu_{e}$ energy spectrum with that predicted from the 3$+$1 neutrino mixing model. 
The event energy spectrum is divided in $N = 6$ bins, as shown in figure~\ref{fig:energy_spectra_no_osc}; $n_{i}$ is the number of observed events in the $i$-th bin. 
The expected number of events in each bin, $\mu_i$, is evaluated using GLoBES~\cite{GLoBES1, GLoBES2}. Detector effects are taken into account by smearing matrices calculated by MC simulation.
The contributions from four neutrino oscillation channels are taken into account, namely $\nu_e(\bar{\nu}_e) \to \nu_e(\bar{\nu}_e)$ and $\nu_{\mu}(\bar{\nu}_{\mu}) \to \nu_e(\bar{\nu}_e)$. The backgrounds arising from $\pi^{0}$ and $\tau \rightarrow e$, defined in section~\ref{sec:nuesearch}, are considered independent of the 3$+$1 mixing model parameters. 
Two corrective and independent factors, $k_{j}$ ($j=1, 2$), are introduced to take into account the overall systematic uncertainties, $\sigma_j$, on the intrinsic $\nu_{e}$ and $\bar{\nu}_{e}$ beam components, on the $\nu_{e}$ detection efficiency and on $\nu_{e}$ and $\bar{\nu}_{e}$ cross sections. The expected number of events is

\begin{equation}
\mu_{i} = \mu_{i}^{0} (1 + k_{j}) \quad \textrm{where} \quad
    \begin{cases}
      j = 1, & \text{if}\ i = 1 \\
      j = 2, & \text{otherwise}
    \end{cases}.
\label{eq:corrected_nexp}
\end{equation}
$\mu_{i}^{0}$ is the expected number of events estimated using the nominal fluxes, cross-sections and detection efficiencies. As already stated, the systematic uncertainties, $\sigma_j$, are conservatively estimated as 20\% for the first energy bin ($E \leq 10$~GeV) and 10\% for $E > 10$~GeV. The profile likelihood ratio was used as test statistic, with the likelihood defined as:

\begin{equation}
-2 \ln{L} = -2 \sum_{i}^{N} \left(n_{i} \ln{\mu_{i}} - N\mu_{i} \right)+ \sum_{j=1}^{2} \frac{k_{j}^{2}}{\sigma_{j}^{2}} + \frac{\left(\Delta m^{2}_{31}-\widehat{\Delta m^{2}_{31}}\right)^{2}}{\sigma^{2}_{\Delta m^{2}_{31}}}.
\label{eq:loglikelihood}
\end{equation}
The last term is a penalty term accounting for the current knowledge on $\Delta m^{2}_{31}$~\cite{PDG2016}, $\widehat{\Delta m^{2}_{31}}$ and $\sigma^{2}_{\Delta m^{2}_{31}}$ are the best fit value and the 1$\sigma$ uncertainty, respectively. The parameters of interest are the squared mass difference $\Delta m^{2}_{41}$ and the effective mixing $\sin^{2}2\theta_{\mu e} = 4|U_{e4}|^{2}|U_{\mu4}|^{2}$. It is worth noting that this definition of the effective mixing allows a direct comparison of this analysis with short-baseline results.
The non-zero value of $\Delta m^{2}_{21}$ is taken into account as well as matter effects assuming a constant Earth crust density estimated with the PREM~\cite{PREM1, PREM2} onion shell model. All other oscillation parameters are treated as nuisance parameters and profiled out. The result is restricted to positive $\Delta m^{2}_{41}$ values  since negative values are disfavoured by results on the sum of neutrino masses from cosmological surveys~\cite{Planck}. The resulting 90\% C.L. exclusion region is shown in figure~\ref{fig:exclusion_plot}. An upper limit on $\sin^2(2\theta_{\mu e}) = 0.021$ is set for $\Delta m_{41}^{2} > 0.1~\textrm{eV}^2$. Moreover, OPERA contributes to limit the effective mixing for low $\Delta m^{2}_{41}$ and excludes $\Delta m_{41}^{2} \gtrsim 4 \times 10^{-3}~\textrm{eV}^{2}$ for maximal mixing. It must be stressed that, for small $\Delta m^{2}_{41}$ values, OPERA is the only experiment having collected data in appearance mode. MINOS and Daya Bay/Bugey-3 have obtained a more stringent exclusion limit in a combined analysis of their disappearance results on, respectively, an accelerator $\nu_{\mu}$ beam and reactors $\bar{\nu}_e$ fluxes.
\section*{Conclusions}
\addcontentsline{toc}{section}{Conclusions}
\label{sec:conclusion}

\par The full OPERA data sample collected during the 2008--2012 CNGS beam runs is used to search for $\nu_{e}$ candidates. Compared to the previous search~\cite{OPERAnue}, results reported here are based on a larger statistics and improved $\nu_{e}$ CC energy resolution. 
Model independent results are reported to allow testing different oscillation hypotheses.
The results are compatible with the no-oscillation hypothesis as well as with the 3 neutrino flavour one. In the latter case, a 90\% C.L. upper limit $\sin^2(2\theta_{13}) < 0.43$ is set. 
For the first time, in this paper, the results are also analysed in a 3+1 model to test the hypothesis of the existence of a sterile neutrino as suggested by several experiments. A 90\% C.L. upper limit $\sin^2(2\theta_{\mu e}) = 0.021$ for $\Delta m^{2}_{41} \gtrsim 0.1$~eV$^2$ is set, thus excluding a sizable part of the region hinted by LSND and MiniBooNE experiments.  OPERA is the only appearance experiment excluding neutrino mass difference down to $4 \times 10^{-3}$~eV$^2$.

\acknowledgments
We wish to thank  CERN for the successful operation of the CNGS facility and INFN for the continuous support given to the experiment through its LNGS laboratory. We warmly acknowledge funding from our national agencies: Fonds de la Recherche Scientifique-FNRS and Institut Interuniversitaire des Sciences Nucleaires for Belgium; MoSES for Croatia; CNRS and IN2P3 for France; BMBF for Germany; INFN for Italy; JSPS, MEXT, the QFPU-Global COE program of Nagoya University, and Promotion and Mutual Aid Corporation for Private Schools of Japan for Japan; SNF, the University of Bern and ETH Zurich for Switzerland; the Russian Foundation for Basic Research (Grant No. 12-02-12142 ofim), the Programs of the Presidium of the Russian Academy of Sciences (Neutrino Physics and Experimental and Theoretical Researches of Fundamental Interactions), and the Ministry of Education and Science of the Russian Federation for Russia, the Basic Science Research Program through the National Research Foundation of Korea (NRF) funded by the Ministry of Science, ICT and Future Planning (Grant No. NRF-2015R1A2A 2A01004532) for Korea; and TUBITAK, the Scientific and Technological Research Council of Turkey for Turkey (Grant No. 108T324). We thank JINR Association of Young Scientists and Specialists for partial support of this work (Grant No. 16-202-02). We thank the IN2P3 Computing Centre (CCIN2P3) for providing computing resources. 


\begin{thebibliography}{99}



\bibitem{OPERAproposal} 
OPERA collaboration, \emph{An appearance experiment to search for $\nu_{\mu}\leftrightarrow \nu_{\tau}$ oscillations in the CNGS beam. Experimental proposal}, SPSC-2000-028, CERN-SPSC-P-318, LNGS-P25-00

\bibitem{OPERA2009}
OPERA collaboration, \emph{The OPERA experiment in the CERN to Gran Sasso neutrino beam}, \emph{JINST} {\bf 4} (2009) P04018

\bibitem{CNGSbeam1}  G.~Acquistapace et al., \emph{The CERN neutrino beam to Gran Sasso (NGS)}, CERN-98-02, INFN/AE-98/05;
R. Bailey et al., \emph{The CERN neutrino beam to Gran Sasso (NGS). Addendum to report CERN-98-02, INFN/AE-98/05}, CERN-SL/99-034(DI), INFN/AE-99/05

\bibitem{tau5Sigma}
OPERA collaboration, \emph{Discovery of $\tau$ neutrino appearance in the CNGS neutrino beam with the OPERA experiment}, \emph{Phys. Rev. Lett.} {\bf 115} (2015)  121802 [arXiv:1507.01417 [hep-ex]]

\bibitem{OPERAnue}
OPERA collaboration, \emph{Search for $\nu_{\mu} \to \nu_e$ oscillations with the OPERA experiment in the CNGS beam}, \emph{JHEP} {\bf 07} (2013) 004 [arXiv:1303.3953 [hep-ex]]

\bibitem{LSND}
LSND collaboration, \emph{Evidence for neutrino oscillations from the observation of $\bar{\nu}_e$ appearance in a $\bar{\nu}_{\mu}$ beam}, \emph{Phys. Rev. D} {\bf 64} (2001) 112007 [arXiv:hep-ex/0104049]

\bibitem{MiniBooNE}
MiniBooNE collaboration, \emph{A combined $\nu_{\mu} \to \nu_e$  and $\bar{\nu}_{\mu} \to \bar{\nu}_e$ oscillation analysis of the MiniBooNE excesses}, FERMILAB-PUB-12-394-AD-PPD, LA-UR-12-23041, [arXiv:1207.4809 [hep-ex]]

\bibitem{reactoranomaly}  G.~Mention, M.~Fechner, T.~Lasserre, T.~A.~Mueller, D.~Lhuillier, M.~Cribier and A.~Letourneau, \emph{The Reactor Antineutrino Anomaly}, \emph{Phys.\ Rev.\ D} {\bf83} (2011) 073006 [arXiv:1101.2755 [hep-ex]]

\bibitem{GALLEX}  F.~Kaether, W.~Hampel, G.~Heusser, J.~Kiko and T.~Kirsten, \emph{Reanalysis of the GALLEX solar neutrino flux and source experiments}, \emph{Phys.\ Lett.\ B} {\bf 685} (2010) 47 [arXiv:1001.2731 [hep-ex]]

\bibitem{SAGE}  J.~N.~Abdurashitov et al., \emph{Measurement of the response of a Ga solar neutrino experiment to neutrinos from an Ar-37 source}, \emph{Phys.\ Rev.\ C} {\bf73} (2006) 045805 [arXiv:nucl-ex/0512041]

\bibitem{Niu:1971xu}  K.~Niu, E.~Mikumo and Y.~Maeda, \emph{A Possible Decay in Flight of a New Type Particle}, \textit{Prog.\ Theor.\ Phys.\ } \textbf{46} (1971) 1644

\bibitem{OPERAspec}
OPERA collaboration, \emph{Determination of the muon charge sign with the dipolar spectrometers of the OPERA experiment}, \emph{JINST} {\bf 11} (2016) 007, P07022 [arXiv:1404.5933 [physics.ins-det]]

\bibitem{OPERA_BF}  Yu.~A. Gornushkin, S.~G. Dmitrievsky, A.~V. Chukanov, \emph{Locating of the neutrino interaction vertex with the help of electronic detectors in the OPERA experiment}, \emph{Phys.\ Part.\ Nucl.\ Lett.\ } {\bf12} (2015) pg.89

\bibitem{OPERA1stTau}
OPERA collaboration, \emph{Observation of a first $\nu_\tau$ candidate in the OPERA experiment in the CNGS beam}, \emph{Phys.\ Lett.\ B} {\bf691} (2010) pg.138 [arXiv:1006.1623 [hep-ex]]


\bibitem{OPERAds}
OPERA collaboration, \emph{Procedure for short-lived particle detection in the OPERA experiment and its application to charm decays}, \emph{Eur.\ Phys.\ J.\ C} {\bf 74} (2014) 8, 2986 [arXiv:1404.4357 [hep-ex]]

\bibitem{OPERA3rdTau}
OPERA collaboration, \emph{New results on $\nu_{\mu} \rightarrow \nu_{\tau}$ appearance with the OPERA experiment in the CNGS beam}, \emph{JHEP} {\bf 11} (2013) 036 [arXiv:1308.2553 [hep-ex]]

\bibitem{OPCARAC}   A. Bertolin and N.~T.~Tran., \emph{OpCarac: an algorithm for the classification of the neutrino interactions recorded by the OPERA experiment}, OPERA public note 100 (2009)

\bibitem{Fluka1}  T.~T. Böhlen., F. Cerutti, M.~P.~W.~Chin, A.~Fasso, A.~Ferrari, P.~G.~Ortega, A.~Mairani, P.~R.~Sala, G.~Smirnov and V.~Vlachoudis, \emph{The FLUKA Code: Developments and Challenges for High Energy and Medical Applications}, \emph{Nucl.\ Data Sheets} {\bf120} (2014) 211

\bibitem{Fluka2}  A.~Ferrari, P.~R.~Sala, A.~Fasso and J.~Ranft, \emph{FLUKA: A multi-particle transport code}, CERN-2005-010, SLAC-R-773, INFN-TC-05-11

\bibitem{CNGSbeam2} 
 CNGS web site, \emph{http://proj-cngs.web.cern.ch/proj-cngs}

\bibitem{GENIE1} C.~Andreopoulos et al., \emph{The GENIE Neutrino Monte Carlo Generator}, \emph{Nucl.\ Instrum.\ Meth.\ A} {\bf 614} (2010) 87-104 [arXiv:0905.2517 [hep-ex]]

\bibitem{GENIE2} C.~Andreopoulos et al., \emph{The GENIE Neutrino Monte Carlo Generator: Physics and User Manual}, [arXiv:1510.05494 [hep-ex]] 

\bibitem{PDG2016} C.~Patrignani et al., \emph{Particle Data Group}, \emph{Chin.\ Phys.\ C} {\bf 40} (2016) 100001 

\bibitem{EM_Longo1975} E.~Longo and I.~Sestili, \emph{Monte Carlo Calculation of Photon Initiated Electromagnetic Showers in Lead Glass}, \emph{Nucl.\ Instrum.\ Meth.} {\bf 128} (1975) 283 [\emph{Erratum ibid} {\bf 135} (1976) 587]

\bibitem{EM_Param2000} G.~Grindhammer and S.~Peters, \emph{The Parameterized Simulation of Electromagnetic Shower in Homogeneous and Sampling Calorimeters}, [arXiv:000102 [hep-ex]]

\bibitem{Leroy2000} C.~Leroy and P.G.~Rancoita, \emph{Physics of cascading shower generation and propagation in matter: principles of high-energy, ultrahigh-energy and compensating calorimetry}, \emph{Rept.\ Prog.\ Phys.} {\bf 63} (2000) 505

\bibitem{ErecProceeding}  S.~Zemskova, \emph{$\nu_{\mu} \to \nu_e$ oscillations search in the OPERA experiment}, \emph{Phys.\ Part.\ Nucl.} {\bf 47} (2016) 1003

\bibitem{FC}  G.~J. Feldman, R.~D. Cousins, \emph{Unified approach to the classical statistical analysis of small signals}, \emph{Phys.\ Rev.\ D} {\bf 57} (1998) 3873 [arXiv:physics/9711021 [physics.data-an]]

\bibitem{Bayesian}  G.~Cowan, \emph{Statistical Data Analysis}, Clarendon Press, Oxford U.K. (1998)

\bibitem{NOMAD} NOMAD collaboration, \emph{Search for $\nu_{\mu} \to \nu_e$ oscillations in the NOMAD experiment}, \emph{Phys.\ Lett.\ B} {\bf 570} (2003) 19 [hep-ex/0306037]

\bibitem{KARMEN} KARMEN collaboration, \emph{Upper limits for neutrino oscillations $\bar{\nu}_{\mu} \to \bar{\nu}_e$ from muon decay at rest}, \emph{Phys.\ Rev.\ D} {\bf 65} (2002) 112001 [hep-ex/0203021]

\bibitem{MINOSDB} Daya Bay and MINOS collaborations, \emph{Limits on Active to Sterile Neutrino Oscillations from Disappearance Searches in the MINOS, Daya Bay, and Bugey-3 Experiments}, \emph{Phys.\ Rev.\ Lett.} {\bf 117} (2016) no.15,  151801 [arXiv:1607.01177 [hep-ex]]

\bibitem{GLoBES1}  P.~Huber, M.~Lindner and W.~Winter, \emph{Simulation of long baseline neutrino oscillation experiments with GLoBES}, \emph{Comput.\ Phys.\ Commun.\ } {\bf 167} (2005) 195 [arXiv:hep-ph/0407333]

\bibitem{GLoBES2}  P. Huber, J. Kopp, M. Lindner, M. Rolinec, and W. Winter, \emph{New features in the simulation of neutrino oscillation experiments with GLoBES 3.0}, \emph{Comput.\ Phys.\ Commun.} {\bf 177} (2007) 432 [arXiv:hep-ph/0701187]

\bibitem{PREM1}  A.~M. Dziewonski and D.~L. Anderson, \emph{Preliminary reference earth model}, \textit{Phys.\ Earth Planet.\ Interiors} {\bf 25} (1981) 297

\bibitem{PREM2}  F.~D.~Stacey and P.~M.~Davis, \emph{Physics of the Earth}, 4th ed., Cambridge University Press, Cambridge U.K. (2008)

\bibitem{Planck}
Planck collaboration, \emph{Planck 2015 results. XIII. Cosmological parameters}, \emph{A\&A} {\bf 594}, A13 (2016) [arXiv:1502.01589 [astro-ph.CO]]

\end{thebibliography}
\end{document}